\documentclass[fleqn,usenatbib]{mnras}

\usepackage{newtxtext,newtxmath}

\usepackage[T1]{fontenc}

\DeclareRobustCommand{\VAN}[3]{#2}
\let\VANthebibliography\thebibliography
\def\thebibliography{\DeclareRobustCommand{\VAN}[3]{##3}\VANthebibliography}

\usepackage{graphicx}	
\usepackage{amsmath}	

\title[Self-interacting superfluid DM droplets]{Self-interacting superfluid dark matter droplets}

\author[V. Delgado and A. Mu\~noz Mateo]{
Vicente Delgado$^{1}$
and Antonio Mu\~noz Mateo$^{1}$
\\
$^{1}$Departamento de F\'{i}sica, Facultad de Ciencias, Universidad de La Laguna, E--38200 La Laguna, Tenerife, Spain
}

\date{Accepted XXX. Received YYY; in original form ZZZ}

\pubyear{2022}

\begin{document}
\label{firstpage}
\pagerange{\pageref{firstpage}--\pageref{lastpage}}
\maketitle

\begin{abstract}
We assume dark matter to be a cosmological self-gravitating Bose--Einstein condensate of 
non-relativistic ultralight scalar particles with competing gravitational and repulsive 
contact interactions and investigate the observational implications of such model.
The system is unstable to the formation of stationary self-bound structures that minimize 
the energy functional. 
These \textit{cosmological superfluid droplets}, which are the smallest possible gravitationally 
bound dark matter structures, exhibit a universal mass profile and a corresponding
universal rotation curve.
Assuming a hierarchical structure formation scenario where granular dark matter haloes 
grow around these primordial stationary droplets, the model predicts cored haloes with 
rotation curves that obey a single universal equation in the inner region 
($r\,\lesssim\,1$\,kpc).
A simultaneous fit to a selection of galaxies from the SPARC database chosen with the 
sole criterion of being strongly dark matter dominated even within the innermost region, 
indicates that the observational data are consistent with the presence of a Bose--Einstein 
condensate of ultralight scalar particles of mass $m \simeq 2.2 \times 10^{-22}$\,eV\,c$^{-2}$ 
and repulsive self-interactions characterized by a scattering length 
$a_s \simeq 7.8 \times 10^{-77}$\,m. Such small self-interactions have profound 
consequences on cosmological scales. They induce a natural minimum scale length for the 
size of dark matter structures that makes all cores similar in length ($\sim 1$\,kpc) and 
contributes to lower their central densities.
\end{abstract}




\section{Introduction}

Despite the great success of the Lambda Cold Dark Matter ($\Lambda$CDM)
model in describing the large-scale structure of the Universe, on smaller scales
($\lesssim 10$\,kpc) a number of predictions of this model seem to be in clear disagreement 
with observational evidence \citep{Weinberg2015, Bullock2017}.
In particular, CDM N-body simulations yield an overabundance of unobserved 
small-scale structure \citep{Kauffmann1993, Klypin1999, Moore1999}
and predict dark matter haloes with dense cuspy density profiles at 
their centres \citep{Navarro1997}, while observations in dwarf 
galaxies favour shallow cored halo profiles 
\citep{Flores1994, Moore1994, Boylan2011}.
Observations also indicate that these dark-matter dominated galaxies have a minimum characteristic 
scale length as well as remarkably similar central densities over a wide range of luminosities, with 
a mass content within the innermost 300 pc of 
$\sim 10^7$\,M$_{\sun}$ 
\citep{Gilmore2007, Strigari2008}.
These results and, in general, the observed properties of the mass profiles at the centres of 
the haloes are difficult to explain within the standard pressureless CDM model. 

While the problems of the CDM model can be alleviated in part by introducing specific baryonic 
physics \citep{Weinberg2015,Bullock2017}, in this work we are interested on a different approach 
that considers dark matter as a Bose-condensed scalar field. 
According to this proposal, which has a long history 
\citep{Baldeschi1983,Khlopov1985,Colpi1986,Sin1994,Lee1996,Guzman2000,Sahni2000,%
Hu2000,Matos2001,Silverman2002,Robles2013,Berezhiani2015,Li2017,%
Deng2018,Chavanis2019a,Chavanis2019b,Berezhiani2021},
a system of non-relativistic scalar particles characterized by a cosmologically large de Broglie wavelength 
occupy their lowest energy state forming a Bose--Einstein condensate (BEC) that can be described in terms of a 
coherent wave function $\Psi$.
Such a system behaves as a quantum fluid whose pressure induces a non-zero Jeans scale that can help solve the 
small-scale problems of the CDM model by suppressing the excess of substructure and producing cored density 
profiles \citep{Sahni2000,Hu2000}.
Indeed, a BEC has non-zero pressure even at zero temperature. In particular,
the quantum pressure, which has its origin in the uncertainty principle, always contributes. 
This component is a direct manifestation of the zero-point motions of the constituent particles.
On the other hand, BECs with local interparticle interactions characterized by a coupling constant $g$ 
have a second component of the form $P=g|\Psi|^2/2$.
For condensates with sufficiently large repulsive ($g>0$) interaction strengths or particle 
densities $|\Psi|^2$, this contribution may be dominant.
When this occurs and the quantum pressure becomes negligible against interparticle interactions
the system enters the Thomas--Fermi (TF) regime. 
Self-gravitating BECs with negligible \citep{Hu2000}, dominant \citep{Goodman2000,Bohmer2007}
and arbitrary short-range interparticle interactions \citep{Chavanis2011a,Chavanis2011b,Lora2012} have been
considered as possible candidates for dark matter haloes.

More recently, motivated by high-resolution cosmological simulations based on the Schr\"{o}dinger--Poisson 
equations \citep{Schive2014a,Schive2014b,Mocz2019}, attention has focused mainly on axionlike dark matter 
particles with negligible (attractive) self-interactions 
\citep{Marsh2014,Hlozek2015,Guth2015,Marsh2016,Hui2017,%
Bar2018,Bar2019,Robles2019,BarOr2019,Safarzadeh2020,%
Ferreira2021,Li2021,Hui2021,Hayashi2021,Chen2021}.
These simulations have confirmed that an axionlike particle with $m \approx 10^{-22}$\,eV\,c$^{-2}$ can help solve
the small-scale problems of the CDM model while maintaining its successful large-scale predictions. 
In particular, they show a clear suppression of substructure below 
$\sim 10^{8}$\,M$_{\sun}$ 
and predict cored haloes
with a granular structure dominated by a central \textit{soliton} (whose radius scales inversely with its mass)
surrounded by an outer density profile that mimics a Navarro--Frenk--White (NFW) profile \citep{Navarro1997}. 

Other numerical estimates based on axionlike particles with no self-interactions lead to values for the mass
of dark matter particles that, in general, are consistent with the above result 
\citep{Lora2012,Marsh2015,Calabrese2016,Schive2016,Chen2017,Urena2017,Church2019}. 
However, the suppression of small-scale structure that such a light mass would induce seem to be in
tension with observations of the Lyman-$\alpha$ forest, which rule out boson masses 
$m \lesssim 10^{-21}$\,eV\,c$^{-2}$ \citep{Hui2017,Irsic2017,Armengaud2017,Rogers2021}.
Nevertheless, this issue is still under debate. There is some consensus that a value of 
$m \sim 10^{-22}$\,eV\,c$^{-2}$ provides the most relevant cosmological consequences and it has been suggested 
that perhaps more sophisticated models of reionization could alleviate this tension \citep{Hui2017,Broadhurst2020,Pozo2021}.

In this work, we assume dark matter to be a self-gravitating BEC consisting of non-relativistic ultralight 
scalar particles with competing gravitational and short-range repulsive interparticle interactions and 
investigate the observational implications of such model.
Self-gravitating BECs with arbitrary short-range repulsive or attractive self-interactions
has been thoroughly studied by \citet{Chavanis2011a,Chavanis2011b,Chavanis2021}
\citep[for other works concerning repulsive self-interactions, see also][]{%
Robles2012,Suarez2017,Bernal2018,Padilla2021,Dawoodbhoy2021,Shapiro2022}.

\begin{figure}
\includegraphics[width=\columnwidth]{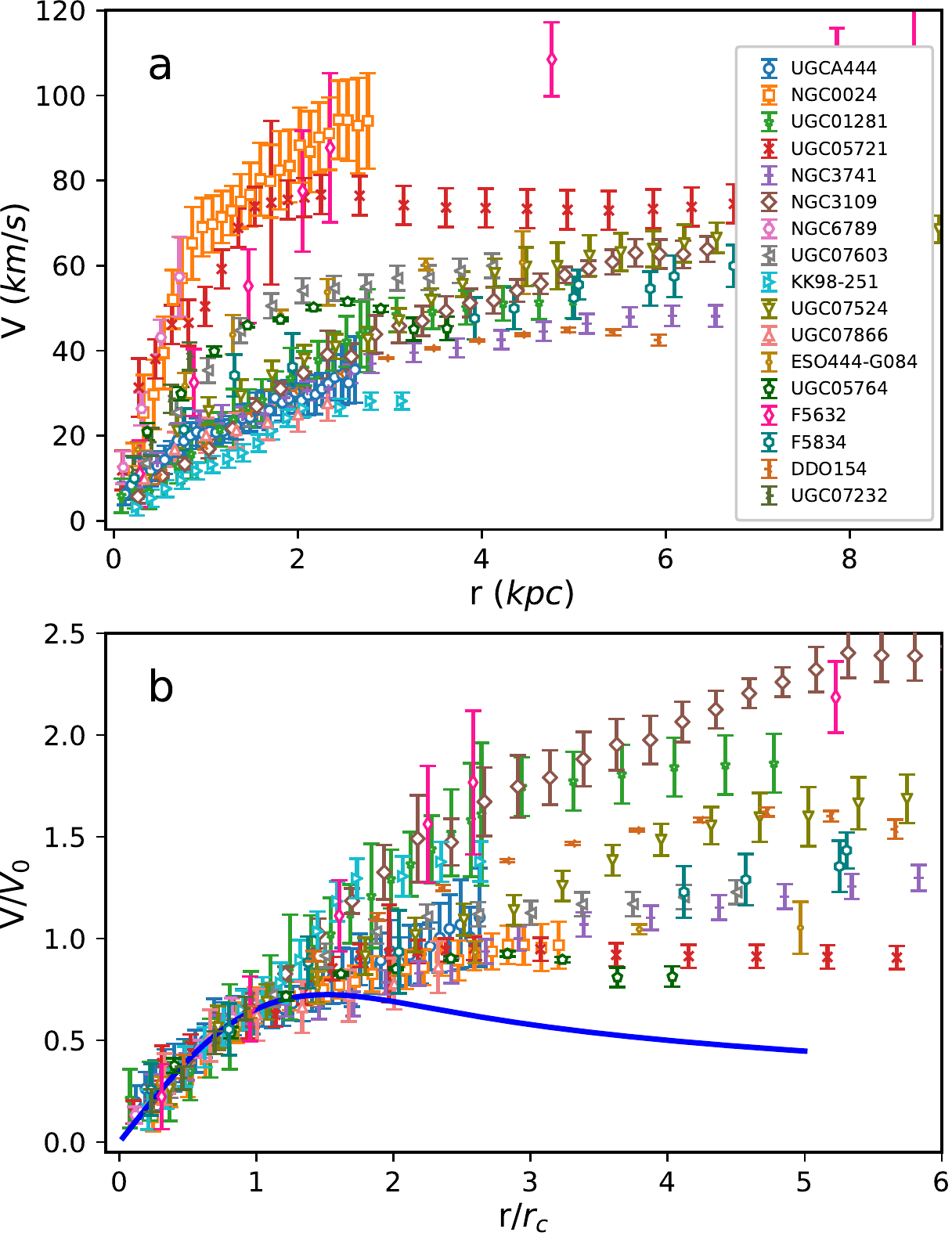}
\caption{Symbols correspond to the observed dark matter rotation curves of the
most dark matter dominated galaxies in the SPARC database. In the top panel
velocities and radii are in physical units. 
In the bottom panel velocities are in units of $V_0=\sqrt{GM/r_c}$ and radii
in units of $r_c$. This figure shows that within the range of applicability
of the theoretical model (for $r \lesssim r_c \sim 1$\,kpc) the observational
data can be well fitted by the universal rotation curve (\ref{velo}) 
(solid blue line). }%
\label{Fig1}%
\end{figure}

The system is unstable to the formation of stationary self-bound structures that minimize the 
energy functional. These are the smallest possible gravitationally bound dark matter structures, which we 
will refer to as \textit{cosmological superfluid droplets}. 
Guided by recent numerical simulations \citep{Schive2014a,Schive2014b,Schwabe2016,Veltmaat2016,Mocz2017}, 
we assume a hierarchical structure formation with cored dark matter haloes growing around these primordial 
stationary droplets. With this assumption the model predicts rotation curves that obey a single universal 
equation in the inner region.  
A simultaneous fit to a selection of galaxies from the SPARC database \citep{Lelli2016,Li2020} chosen with 
the sole criterion of being strongly dark matter dominated even in the innermost region indicates that 
the observational data are consistent with the presence of a BEC of ultralight scalar particles of mass 
$m \simeq 2.2 \times 10^{-22}$\,eV\,c$^{-2}$ and repulsive contact interactions characterized by a s-wave 
scattering length $a_s \simeq 7.8 \times 10^{-77}$\,m.
This result follows from Fig. \ref{Fig1}, which will be considered in more detail below.
While most of the recent theoretical and numerical works have focused on axionlike dark matter particles 
with negligible (attractive) self-interactions ($a_s=0$), the effects of the above small repulsive 
self-interactions turn out to be essential on cosmological scales. 
They induce a natural minimum scale length for the size of (non-linear) dark matter structures that, in 
particular, makes all superfluid droplets have a similar length ($\sim 1$\,kpc) and contributes to lower 
the central densities of massive haloes, which may help to solve the small-scale problems of the CDM model.


\section{Self-interacting BEC dark matter}

A dark matter particle of mass $m \sim 10^{-22}$\,eV\,c$^{-2}$ and velocity $v\sim 100$\,km\,s$^{-1}$ 
\citep{Herzog2018} 
has a de Broglie wavelength $\lambda_\mathrm{dB} =h/m v \sim 1.2$\,kpc. This wavelength greatly exceeds 
the interparticle separation $\bar n^{-1/3}$ (where $\bar n$ is the average particle-number density), 
which for typical galactic matter densities $\rho\sim 1$\,GeV\,c$^{-2}$\,cm$^{-3}$ is of the order 
$\bar n^{-1/3} \sim 10^{-10}$\,cm. Under these conditions, dark matter is deep in the quantum degenerate 
regime ($\lambda_\mathrm{dB}\,\bar n^{1/3}\gg\,1$) and must be in the form of a BEC.
From the success of the Gross--Pitaevskii equation in reproducing the outcomes of laboratory experiments 
with ultracold gases, it is now a well-established fact that this equation provides an accurate description 
of the dynamics of zero-temperature BECs in the mean field regime 
\citep[see e.g.][]{VDB2006,Huhtamaki2006}. 
For non-relativistic self-gravitating BECs, it takes the form
\begin{equation}
 i\hbar\frac{\partial}{\partial t}\Psi =
 \left(-\frac{\hbar^2}{2m} \,\nabla^2+g\,|\Psi|^2 + m\,\Phi_\textsc{g} \right)\Psi ,
 \label{tdgpe}
\end{equation}
where the strength of the short-range interparticle interactions, $g=4\pi \hbar^2 a_s / m$, is determined 
by the s-wave scattering length $a_s$, and $\Phi_\textsc{g}$ is the gravitational potential that satisfies 
the Poisson equation:
\begin{equation}
\nabla^2\Phi_\textsc{g}=4\pi G \rho,
 \label{poisson}
\end{equation}
where $\rho(\mathbf{r},t)\equiv m|\Psi(\mathbf{r},t)|^2$ is the condensate mass density. The ground-state 
solutions of the dark matter field (cosmological quantum droplets) are localized, self-bound stationary 
solutions of the coupled Gross--Pitaevskii--Poisson (GPP) equations above that can be found by writing the 
order parameter in the form $\Psi(\mathbf{r},t)=\exp(-i\mu_c \,t/\hbar)\,\psi(\mathbf{r})$, where $\mu_c$ 
is the chemical potential and $\psi$ satisfies the time-independent equation
\begin{equation}
\mu_c \psi =
 \left(-\frac{\hbar^2}{2m} \,\nabla^2 + g n + m \Phi_\textsc{g}\right) \psi ,
 \label{gpe}
\end{equation}
where $n(\mathbf{r},t)\equiv |\Psi(\mathbf{r},t)|^2$ is the particle-number density of the condensate.
Solving equation~(\ref{gpe}) is equivalent to finding the critical points of the energy functional
\begin{equation}
E[\psi] = \int d\mathbf{r} \left[ \frac{\hbar^2}{2m}\left|{\nabla 
\psi}\right|^2 + \frac{1}{2}{g n^2}  + \frac{1}{2}{m n}\Phi_\textsc{g}\right].
\label{energy}
\end{equation}
Depending on the total mass $M=m\int d\mathbf{r}|\psi|^2$ and the interaction strength $g$, the ground-state 
isotropic dark matter droplets can evolve from (bright) matter-wave solitons to TF-like quantum droplets, 
with a smooth crossover between the two regimes. In the former case, the system is stabilized by a precise 
balance between gravitational interactions and quantum pressure, while in the latter the gravitational 
interaction is counterbalanced by repulsive interparticle interactions.
These solutions are well described by a variational Gaussian ansatz \citep{Chavanis2011a}:
\begin{equation}
\psi(\mathbf{r})=\sqrt{\frac{M/m}{(\sqrt{\pi}r_c)^3}}
\exp{\left(-\frac{r^2}{2r_c^2} \right)},
\label{Gaussian}
\end{equation}
where $r_c$ is a characteristic radius to be determined by minimizing the energy functional (\ref{energy}).

Substituting (\ref{Gaussian}) in (\ref{energy}), the total energy as a function of the variational 
parameter $r_c$ reads
\begin{equation}
E= \frac{3\hbar^2 M}{4 m^2  r_c^2}+ 
\frac{g M^2}{\sqrt{32\pi^3} m^2 r_c^3}-\frac{G M^2}{\sqrt{2\pi}r_c},
\label{Gaussian_energy}
\end{equation}
where the terms on the right hand side are the contributions from the kinetic $E_k$, contact 
self-interaction $E_\mathrm{int}$, and gravitational $E_\textsc{g}$ energies, respectively.
Minimizing equation~(\ref{Gaussian_energy}) with respect to $r_c$, one obtains the virial relation
$2E_{k}+3E_\mathrm{int}+E_\textsc{g}=0$, which can be used to eliminate the contact-interaction term and 
rewrite the energy  $E=(E_k+2E_\textsc{g})/3$ of the dark matter droplet as 
\begin{equation}
E= \frac{\hbar^2 M}{4 m^2 r_c^2}-\sqrt{\frac{2}{9\pi}}\frac{G M^2}{r_c}. 
\label{min_Eb}
\end{equation}
In order for this energy to be negative the radius $r_c$ must satisfy 
$r_c\geq\sqrt{9\pi/32}\,\xi_\textsc{g} \simeq \xi_\textsc{g}$, where
\begin{equation}
\xi_\textsc{g}=\frac{\hbar^2}{G M m^2}.
\label{GCoherence_length}
\end{equation}
The equality, $r_c \simeq \xi_\textsc{g}$, occurs for gravitational solitons. Indeed, as can be seen from 
equation~(\ref{Gaussian_energy}), dark matter droplets with $r_c$ of the order of $\xi_\textsc{g}$ can fully 
balance the gravitational interaction with solely their (zero-point) kinetic energy.
By analogy with atomic BECs we call $\xi_\textsc{g}$ {\it gravitational coherence length}.
For $r_c \gg \xi_\textsc{g}$ the contribution of the kinetic energy $E_{k}$ becomes always negligible against 
$E_\textsc{g}$, so that dark matter solitons must necessarily have a radius of the order of $\xi_\textsc{g}$. 

From equation~(\ref{GCoherence_length}) it is clear that the physical size of these solitons depends on their 
mass and scales as $M^{-1}$. 
In the absence of short-range interactions ($a_s=0$) this would allow the 
existence of small-sized solitons with a very large mass and density (which scales as $M^{4}$).
Repulsive contact interactions ($a_s>0$) has the desirable effect of introducing an upper limit 
${\mathfrak{M}}$ for the mass of a soliton
\begin{equation}
\mathfrak{M}=\hbar  \sqrt{\frac{3\pi}{8G m a_s}}.
\label{Msol_limit}
\end{equation}
Indeed, it follows from equation~(\ref{Gaussian_energy}) that for dark matter droplets of mass $M$ and radius 
$r_c \simeq \xi_\textsc{g}$, the contribution of the quantum pressure becomes negligible against the 
contact-interaction energy $E_\mathrm{int}$ whenever $M^2 \gg  \mathfrak{M}^2$. 
Repulsive contact self-interactions also induce a lower bound for the length scale of gravitational solitons 
of the order of 
\begin{equation}
\mathfrak{R}= \hbar \sqrt{\frac{a_s}{G m^3 }} \simeq \xi_\textsc{g}(\mathfrak{M}).
\label{Rsol_limit}
\end{equation}
In the limit $M^2 \gg \mathfrak{M}^2$, when the quantum pressure becomes negligible, the dark matter droplets 
lose their solitonic character and evolve into TF-like droplets stabilized only by the repulsive interparticle 
interactions. As follows from equation~(\ref{Gaussian_energy}), these TF-like droplets have a radius of the 
order of $\mathfrak{R}$ (hence, independent of the mass $M$) and, unlike the case of gravitational solitons, 
their densities only grow linearly with $M$. Thus, the parameter $\mathfrak{R}$ determines the length scale of 
the smallest gravitationally bound dark matter structures.
The inclusion in the model of short-range interparticle interactions naturally induce a small-scale cut-off 
dependent only on the microscopic parameters of the condensate.

Minimization of the energy functional gives the radii of the superfluid dark matter droplets 
\citep{Chavanis2011a}, which we conveniently write in the form
\begin{equation}
r_c=\sqrt{9\pi/8} \left( 1 + \sqrt{1+(M / \mathfrak{M})^2} \right) \xi_\textsc{g}.
\label{Gaussian_width}
\end{equation}
In the solitonic regime ($M^2 \ll {\mathfrak{M}}^2$), when local self-interactions can be neglected, 
equation~(\ref{Gaussian_width}) reduces to $r_c\approx \sqrt{9\pi/2} \; \xi_\textsc{g}$, while in the TF regime 
($M^2 \gg {\mathfrak{M}}^2$), when the quantum pressure is negligible \citep{Goodman2000,Bohmer2007}, 
$r_c \approx \sqrt{3}\,\mathfrak{R}$.
The parameter ${\mathfrak{M}}^2$ determines the crossover between the two regimes.

\section{Stationary dark matter droplets. Universal mass profile and rotation curve}

Stationary dark matter droplets are fully characterized by the condensate wave function 
(\ref{Gaussian}) with $r_c$ given by equation~(\ref{Gaussian_width}).
Only two parameters are required to describe the fundamental properties of the Bose--Einstein
condensate, namely, the boson mass $m$ and the s-wave scattering length $a_s$ that determines 
the strength of interparicle interactions. Once these parameters are known, the total mass $M$ 
is sufficient to completely specify a given dark matter droplet.
Remarkably, these superfluid dark matter droplets exhibit a {\it universal mass profile.}
Indeed, the enclosed mass within radius $r$, which is given by
\begin{equation}
M(r)=m\int_0^r d\mathbf{r}|\psi|^2 ,
\label{Encmass}
\end{equation}
when expressed in units of the total mass $M$ and as a function of the dimensionless radius $x=r/r_c$, satisfies 
the following universal mass profile:
\begin{equation}
M_{\textsc{dm}}(x)= \mathrm{erf}(x)-2\pi^{-\frac{1}{2}}\,x \exp{(-x^2)}.
\label{unimass}
\end{equation}

The corresponding rotation curve was obtained by \citet{Chavanis2011a}. In view of equation~(\ref{unimass}),
in this work we find it most convenient for our purposes to rewrite the expression derived by 
\citet{Chavanis2011a} in terms of convenient dimensionless quantities. Indeed,
in units of the characteristic circular velocity $V_0=\sqrt{GM/r_c}$, the rotation curves of the dark matter
droplets as a function of $x=r/r_c$ obey the {\it universal equation}
\begin{equation}
V_{\textsc{dm}}(x)=\sqrt{x^{-1}\,M_{\textsc{dm}}(x)}=\sqrt{x^{-1}\,\mathrm{erf}(x)-2\pi^{-\frac{1}{2}}\,\exp{(-x^2)} }.
\label{velo}
\end{equation}
Thus the present model, which considers dark matter to be a superfluid Bose--Einstein condensate of ultralight
scalar particles with short-range repulsive interparticle interactions, predicts a universal mass profile and
a corresponding universal rotation curve in the inner region ($r\,\lesssim\,r_c)$ 
of dark-matter dominated galactic haloes. 
In what follows, we aim 
to elucidate whether the observational data are consistent with these theoretical predictions.

\section{Observational implications}

We assume a hierarchical structure formation scenario where, under the combined action of gravity and emergent 
effective phase-dependent interactions, granular dark matter haloes grow around the above quantum droplets, 
which we assume survive 
as stationary solutions of the GPP equations at the centres of the haloes.
As already mentioned, this assumption seems to be well supported by recent numerical simulations
\citep{Schive2014a,Schive2014b,Schwabe2016,Veltmaat2016,Mocz2017}.

On the other hand, we will also assume that the model is applicable, to a good approximation, \textit{at least} 
up to a radius $r_{\textsc{f}}=1.2$\,kpc. 
Although the latter assumption is mainly motivated by the need to have sufficient data resolution in the region 
of interest and can, in fact, be relaxed somewhat (see Appendix~\ref{sec:appenA}), it also seems to 
be a reasonable assumption \citep[see e.g.][]{Schive2014a,Hui2017}.

The actual range of applicability of the model corresponds to the region where the theoretical curve is able 
to adequately account for the observational data and, for a given galactic halo, can be written, in general, as 
$r \lesssim \alpha r_c \sim r_c$, where $\alpha$ is an unknown parameter of order unity and $r_c$ is the radius 
of the central quantum droplet. As we will see, for all the galaxies considered in this work $r_c$ turns out to 
be of the order of $1$\,kpc. However, this value is not known a priori, a fact that has determined our fitting 
strategy (see Appendix~\ref{sec:appenA} for details). 

We stress that in this work we have pursued minimal model dependence. In fact, our model relies solely on the 
above two assumptions.
With these hypotheses the model predicts observational rotation curves that must satisfy the universal 
equation~(\ref{velo}) in the inner region of the galactic haloes
\citep[for a previous phenomenological construction of a universal rotation curve see][]{Persic1996,Karukes2016}.
Since the presence of baryonic matter can significantly modify the dark matter distribution in the centre of 
the haloes, for the above assumption to be true baryonic matter should be negligible even in this region.
Indeed, equation~(\ref{velo}) was obtained under the assumption of a negligible contribution of baryonic matter, 
which accordingly was not considered in the coupled GPP equations. If this were not the case, there is no 
reason to believe that equation~(\ref{velo}) remains true, and the corresponding GPP equations would have to be 
solved numerically. Therefore, the application of the above analytical formulation to the innermost regions 
of DM haloes requires galaxies with a negligible contribution of baryonic matter even in this central region.

It is worth mentioning that our model assumes stationary cores located at the centre of the haloes, and 
does not consider the possible random motion around this position [as pointed out for solitonic cores by 
\citet{Chowdhury2021}], which, if effective, could also be a source of additional uncertainty in the 
resulting best-fit parameters.

To confront the model with observations, we use the SPARC database restricting our analysis to high-quality
rotation curves with quality flag $Q=1$ or $2$ \citep{Lelli2016,Li2020}.
This database provides the observed circular velocities $V_\mathrm{obs}$ of 175 late-type galaxies along with 
the contributions from gas $V_\mathrm{gas}$ and stars (disk and bulge) $V_{*}$, so that the dark matter 
contribution can be inferred from the relation
$V_\mathrm{obs}^2 = V_{\textsc{dm}}^2 + V_\mathrm{gas}^2 + \Upsilon_* V_{*}^2$ 
(see \citet{Lelli2016,Li2020} for details). 
We focus on galaxies that are strongly dominated by dark matter even within $r\lesssim 1$\,kpc and 
that are well resolved in this region. This excludes galaxies with bulges, since baryonic matter always has 
a significant contribution in the innermost region of these galaxies. 
We then calculate the dark matter contribution to the rotation curves ($V_{\textsc{dm}}$) of the remaining 
galaxies using the relation above with a constant stellar mass-to-light ratio %
$\Upsilon_* = 0.47$\,M$_{\sun}$/L$_{\sun}$
\citep{McGaugh2014,Lelli2016} and carefully analyse the different 
contributions in the region of interest ($r\lesssim 1$\,kpc). Restricting ourselves to galaxies with a ratio 
$V_{\textsc{dm}}/V_\mathrm{obs}>0.75$ at virtually all points in this region, we are finally left with 17 bulgeless 
galaxies. Figure~\ref{Fig1}a shows the dark matter contribution to the rotation curves of these galaxies. 
As is apparent, the sample covers quite different curve slopes and maximum circular velocities (ranging 
from $\sim$\,30 to 100\,km\,s$^{-1}$), so that, despite observational uncertainties, a simultaneous fit of the 17 
curves in the sample with the predicted universal rotation curve (\ref{velo}) imposes a stringent constraint on 
the parameters of the model. Figure~\ref{Fig1}b shows the result of such a fit. 
As fitting parameters we have used $m$, $a_s$ and the different total 
masses $M_i$ of the dark matter droplets residing at the centre of each galaxy and have implemented a 
Maximum Likelihood fitting method that minimizes the standard $\chi^2$ function 
$ \sum (V_{\textsc{dm}}^\mathrm{obs} - V_{\textsc{dm}}^\mathrm{theo})^2 / (\Delta V_{\textsc{dm}}^\mathrm{obs})^2$,
where $V_{\textsc{dm}}^\mathrm{obs}$ is the dark matter contribution to the observed rotation velocity, 
$\Delta V_{\textsc{dm}}^\mathrm{obs}$ is its observational uncertainty and $V_{\textsc{dm}}^\mathrm{theo}$ is the 
theoretical prediction (\ref{velo}). 


\begin{figure}
\includegraphics[width=\columnwidth]{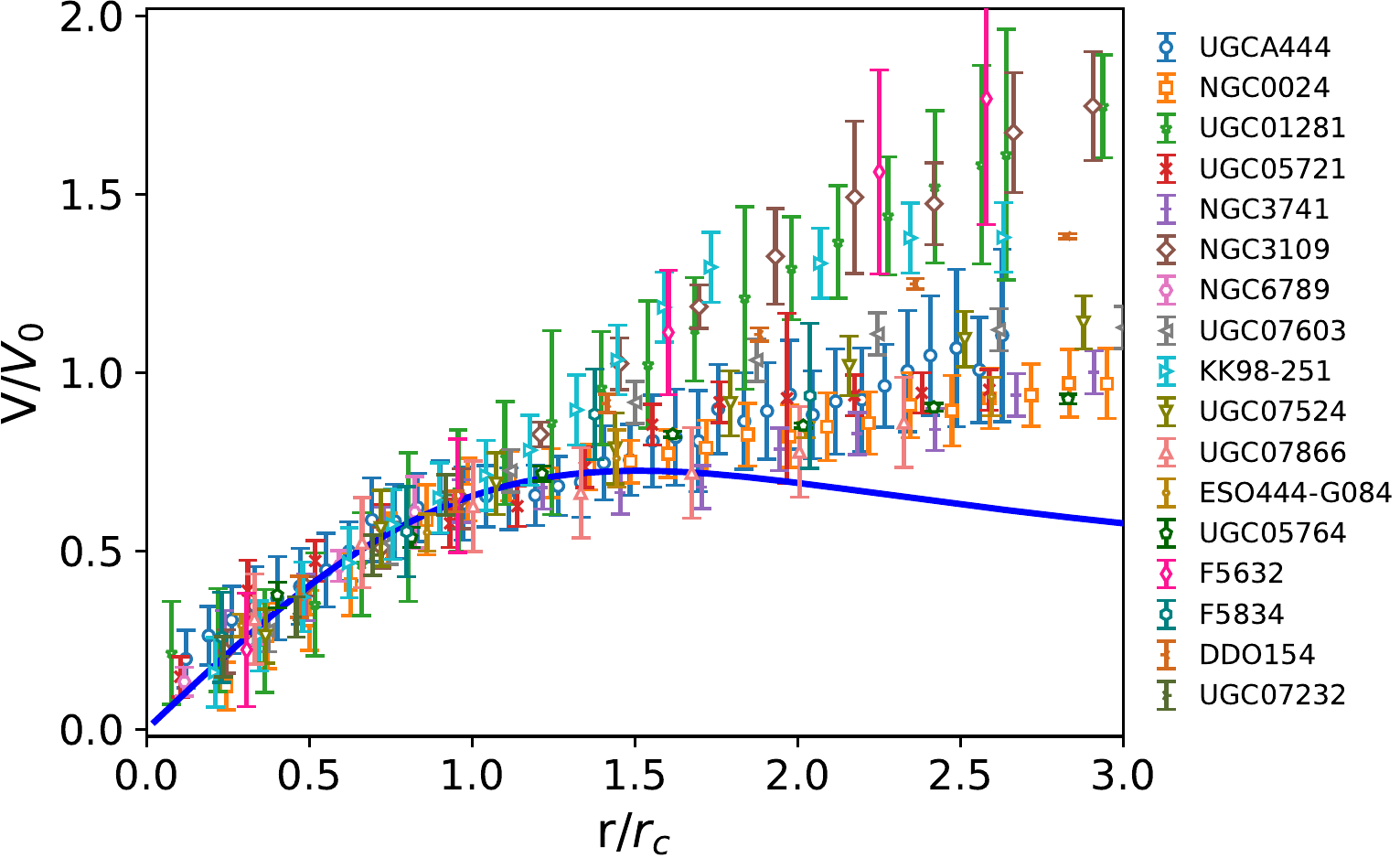}
\caption{This figure is a zoom-in of Fig.~\ref{Fig1}b. 
Symbols correspond to the observed dark matter rotation curves of the
most dark matter dominated galaxies in the SPARC database (see legend).
Velocities are in units of $V_0=\sqrt{GM/r_c}$ and radii in units of $r_c$. 
The figure shows that within the range of applicability of the theoretical 
model (for $r\,\lesssim\,r_c\,\sim\,1$\,kpc) the observational data can be 
well fitted by the universal rotation curve (\ref{velo}) (solid blue curve). }%
\label{Fig2}%
\end{figure}



\begin{figure*}
\includegraphics[width=14.0cm]{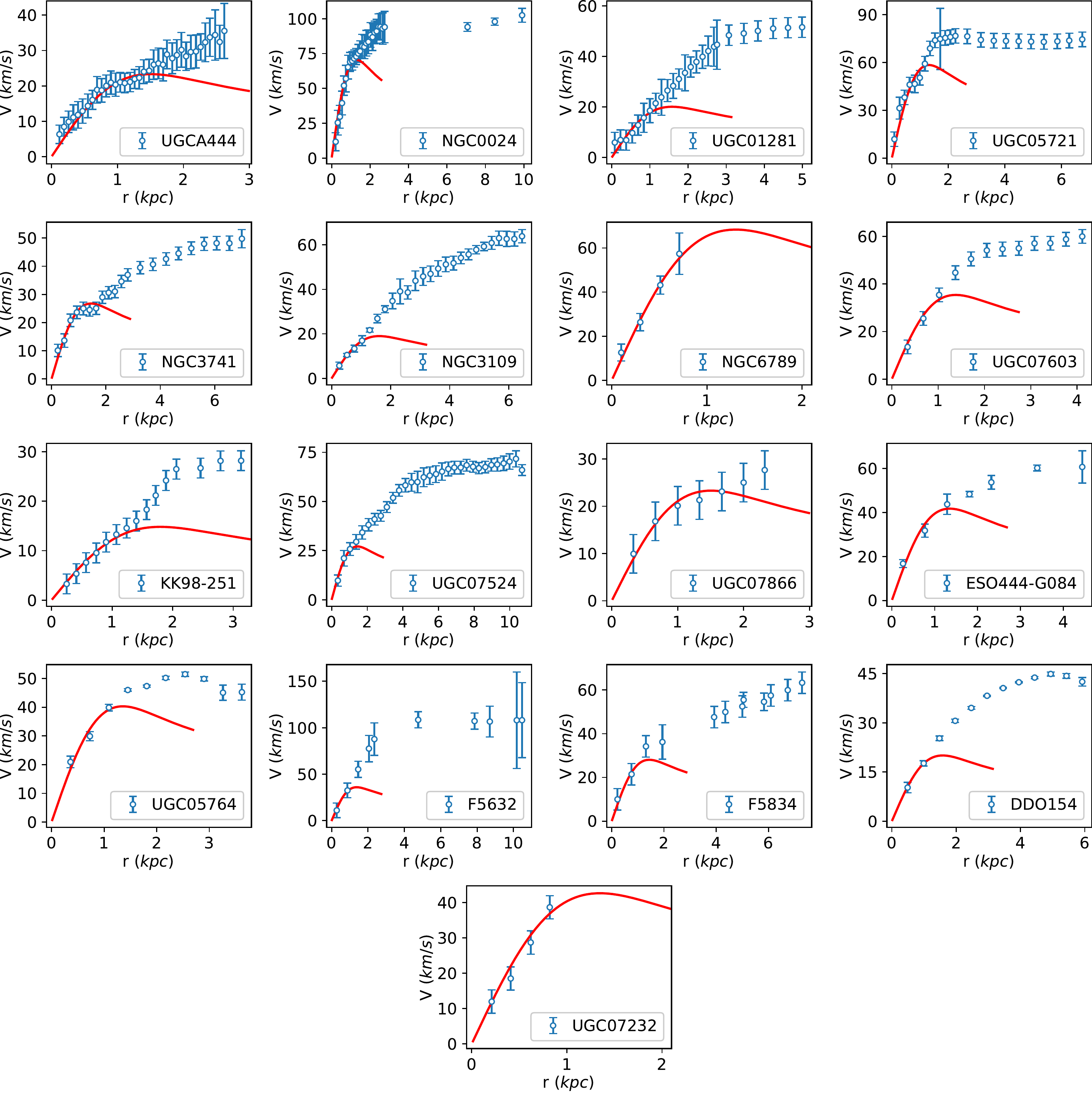}
\caption{Same rotation curves as in Fig.~\ref{Fig2} but in physical units 
(velocities in km\,s$^{-1}$ and radii in kpc). Solid red lines are the fits 
with the universal rotation curve (\ref{velo}). }
\label{Fig3}%
\end{figure*}


As the model is only applicable in the inner region of the haloes, in fitting the observational data we have 
only considered rotation velocities corresponding to radii $r \leq 1.2$\,kpc. On the other hand, since the 
galaxies in the sample have very different spatial resolutions in this region, we have limited the maximum 
number of points to fit per galaxy to 6 to avoid overweighting some galaxies too much.
The best-fit parameters (maximum likelihood) are collected in Table~\ref{Tab1}.
In this table we also show the radius $r_c$ of the halo cores, obtained from equation~(\ref{Gaussian_width}),
as well as the masses of the halo cores $M_c \equiv M(r_c)\simeq 0.43 \, M$, which follow from the definition 
(\ref{Encmass}) of the enclosed mass within radius $r_c$ (see also Fig.~\ref{FigA1}).
Table~\ref{Tab1} also gives the enclosed masses within the innermost 300\,pc, $M_{300}$, and the corresponding 
mean densities, $\rho_{300}$.


\begin{figure}
\includegraphics[width=8.3cm]{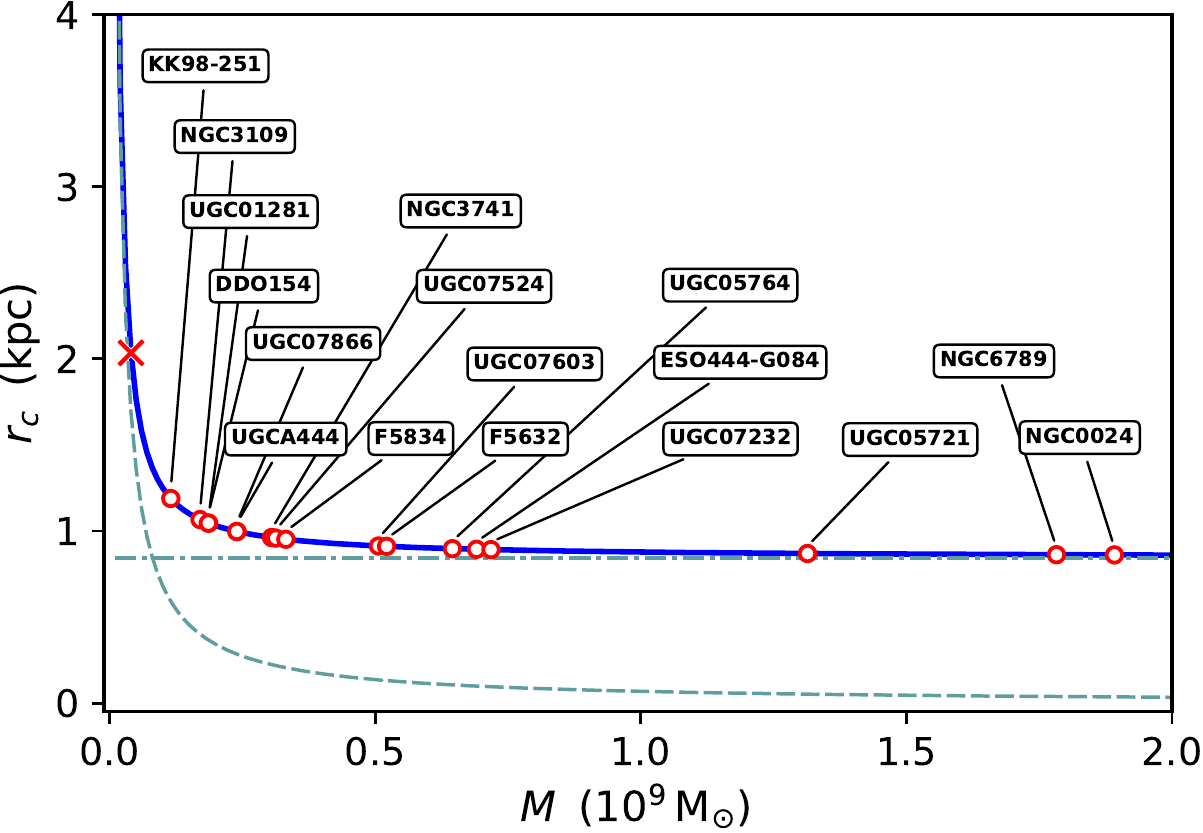}
\vspace{-1mm}
\caption{Location of galaxies (open circles) in the $r_c$--$M$ parameter space. The solid line 
corresponds to equation~(\ref{Gaussian_width}), while the dashed and dot-dashed lines correspond, 
respectively, to the solitonic and TF regimes.
The cross symbol marks the location of a dark matter droplet of mass $\mathfrak{M}$. 
As can be seen, the cores of all galaxies in the sample are either in the TF or in 
the crossover regime, with radii $\sim 1$\,kpc.}%
\label{Fig4}%
\end{figure}



\begin{figure*}
\includegraphics[width=15.5cm]{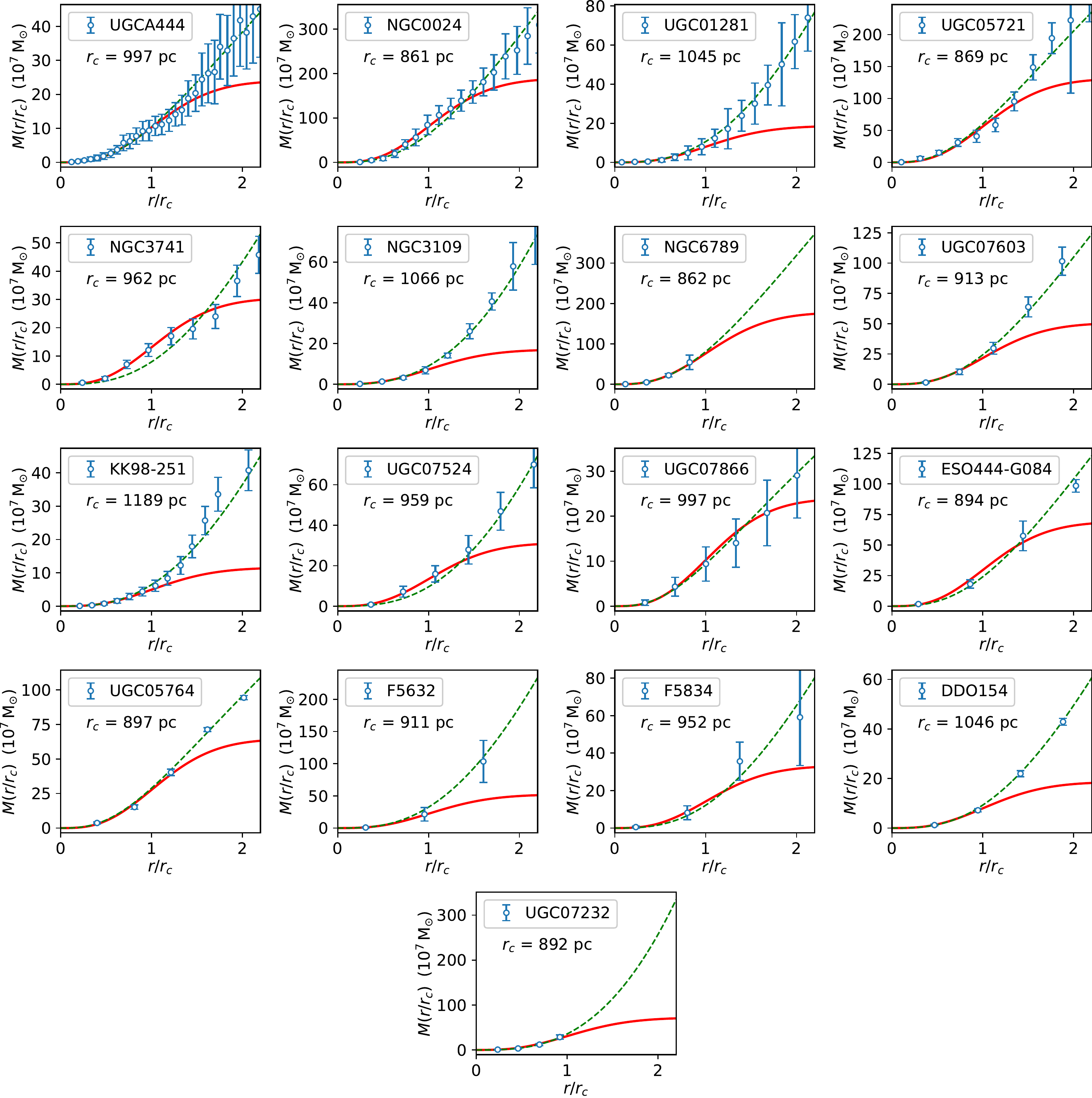}
\caption{Dark matter mass enclosed within radius $r/r_c$. Open symbols correspond to the 
observational data, dashed green lines are fits with the Burkert model and solid red lines are fits 
with the enclosed mass predicted by equation~(\ref{unimass}). 
Note that if we had expressed the enclosed masses in units of $M$ we would have obtained a single universal 
mass profile within 
$r\,\lesssim\,r_c\,\sim\,1$\,kpc.
In fact, this figure contains essentially the same information
as Figs.~\ref{Fig2} and \ref{Fig3}. While the total masses $M$ of the superfluid quantum droplets residing at
the halo centres are, in general, not observable, the above plots show that the mass profiles resulting from 
them are not only consistent with the observed data but, in general, within the range of applicability of the 
model ($r\,\lesssim\,r_c$) fit them better than Burkert's phenomenological model.}
\label{Fig5}%
\end{figure*}


Figure~\ref{Fig1}b shows that within the range of applicability of the theoretical model (for 
$r\,\lesssim\,r_c\,\sim\,1$\,kpc) the observational data can be well fitted by the  
universal rotation curve (\ref{velo}) predicted by the self-interacting BEC model 
(solid blue curve). See also Figs.~\ref{Fig2} and \ref{Fig3}.
Figure~\ref{Fig2} is a zoom-in of Fig.~\ref{Fig1}b while Fig.~\ref{Fig3} shows what the 
universal rotation curve looks like for each individual galaxy in physical units (velocities 
in km\,s$^{-1}$ and radii in kpc). The good agreement between theoretical predictions and
observations indicates that the observational rotation curves are consistent with the 
presence of a superfluid dark matter droplet residing at the centre of galactic haloes and consisting of 
ultralight scalar particles of mass $m \simeq 2.2 \times 10^{-22}$\,eV\,c$^{-2}$ and repulsive contact 
interactions characterized by a s-wave scattering length $a_s \simeq 7.8 \times 10^{-77}$\,m. For these 
parameters, we have
\begin{equation}
\mathfrak{M}=4\times 10^7\,\mathrm{M}_{\sun}, \qquad \mathfrak{R}=487\,\mathrm{pc}.
\label{Param}
\end{equation}

Figure~\ref{Fig4} shows the location of each galaxy (open circles) in the $r_c$--$M$ parameter space. The 
solid line corresponds to equation~(\ref{Gaussian_width}), while the dashed and dot-dashed lines correspond, 
respectively, to the solitonic and TF regimes.
The latter, characterized by a constant radius, $r_c \approx \sqrt{3}\,\mathfrak{R} = 843$\,pc, determines 
the minimum length scale of any dark matter structure. The cross symbol marks the location of a dark matter 
droplet of mass $\mathfrak{M}$. Droplets with mass $M > \mathfrak{M}$ and radius $r_c < r_c(\mathfrak{M}) 
\approx 2$\,kpc are not supported by quantum zero-point motions (quantum pressure) and
cannot be considered to be solitons.
As can be seen, the cores of all galaxies in the sample are either in the TF regime or in the crossover
between the two regimes, with radii $\sim 1$\,kpc, ranging from 861\,pc (for NGC0024) to 1189\,pc (for KK98--251) 
and corresponding masses ranging from 
$1.9\times 10^9\,\mathrm{M}_{\sun}$ to $1.2\times 10^8\,\mathrm{M}_{\sun}$, 
respectively.
Note that those galaxies whose cores are in the TF regime could also be described analytically by an 
alternative formulation to equation~(\ref{Gaussian}) \citep[see e.g.][]{Goodman2000,Bohmer2007}.
However, our formulation provides a richer scenario, as it can account for the smooth transition between the 
solitonic and TF regimes.

In Fig.~\ref{Fig5} we show that, in general, within the range of applicability of the model ($r\,\lesssim\,r_c$),
the universal mass profile implied by the above masses fits the observational data better than Burkert's 
phenomenological model \citep{Burkert1995}.
On the other hand, the mass contained within $r < 300$\,pc varies between $5.6\times 10^7\,\mathrm{M}_{\sun}$ 
(NGC0024) and $1.3\times 10^6\,\mathrm{M}_{\sun}$ (KK98--251) with a mean value of 
$1.6\times 10^7\,\mathrm{M}_{\sun}$ and mean density $0.14\,\mathrm{M}_{\sun} /\mathrm{pc}^3$, results that are 
in good agreement with typical values observed in the most dark matter dominated galaxies in the Local Group 
\citep{Gilmore2007,Strigari2008}.
Note that our results for $m$ and $a_s$ are consistent with those corresponding to what was called BECt model 
by \citet{Chavanis2021}. However, according to our results (see Fig.~\ref{Fig4}), for a given pair $(m,a_s)$ dark
matter droplets can be in different regimes depending on their masses $M$.

\section{Conclusions}

Despite the undeniable success of the standard CDM model on large scales, this model can hardly explain 
the observed mass profiles at the centres of galactic haloes. 
Motivated by the small-scale problems of the CDM model, in this work we have considered dark matter to be 
a quantum fluid in the form of a self-gravitating Bose--Einstein condensate consisting of non-relativistic 
ultralight scalar particles with competing gravitational and repulsive contact interactions characterized
by a s-wave scattering length $a_s \geq 0$. This quantum system, governed by a cosmologically large coherent 
wave function $\Psi$, is unstable to the formation of stationary self-bound cosmological droplets that minimize 
the energy functional and constitute the smallest possible gravitationally bound dark matter structures. 
We have shown that these superfluid dark matter droplets exhibit a universal mass profile and a corresponding 
universal rotation curve.
We have also assumed a hierarchical structure formation scenario with granular dark matter haloes growing 
around these primordial stationary droplets, which survive at the centres of the haloes. With this hypothesis 
the model predicts cored dark matter haloes that must satisfy a universal mass profile and a universal 
rotation curve in the inner region ($r \lesssim r_c$).

To elucidate whether these theoretical predictions are consistent with the observational data,
we have simultaneously fitted the observed rotation curves of the most dark matter dominated galaxies in 
the SPARC database with the predicted universal rotation curve (\ref{velo}) using as free parameters the 
mass $m$ and scattering length $a_s$ of the constituent bosons and the total mass $M$ of the dark matter 
quantum droplets residing at the centre of each halo.

Although there are other works in the literature that fit galaxies from the
SPARC database \citep[see e.g.][]{Bernal2018,Kendall2020}, they are restricted to
axionlike dark matter particles with no self-interactions or are based on
theoretical models that have not allowed to constrain the properties of the
dark matter particle.
On the other hand, much of this work relies on the core-halo relation \citep{Schive2014b}, 
the validity of which has not yet been fully confirmed \citep[see e.g.][]{Kendall2020}.
In contrast, in the present work we have sought minimal model dependence.
Our model relies on only two assumptions: i) dark matter dominated galactic cores are well 
described by a stationary state of the GPP equations and, ii) the model is applicable at 
least up to a radius $r_{\textsc{f}}=1.2$\,kpc.

Our approach 
takes advantage
of the predicted universal character of the mass profiles and corresponding
rotation curves in the innermost region of the most dark matter dominated
haloes to simultaneously fit the observational data of all galaxies in the
sample, thus increasing their statistical significance.
In this way, we have been able to demonstrate remarkable agreement between
observations and theoretical predictions and have obtained precise estimates
for both the mass $m$ and the scattering length $a_s$ of dark matter particles.
To our knowledge, this is the first time that a precise estimate of $a_s$ has
been inferred from observational data (rather than a wide range of possible
values, spanning many orders of magnitude), as well as the first evidence,
based on a motivated theoretical model, for universal properties of the mass 
profiles at the centres of galactic haloes.

Our best-fit results indicate that observational data are consistent with the existence of self-gravitating 
superfluid dark matter droplets in the centres of the haloes that satisfy a universal rotation curve and consist 
of ultralight scalar particles with very weak repulsive local self-interactions. Despite the small value of the 
scattering length, $a_s \sim 10^{-76}$ m, these self-interactions have profound consequences on cosmological 
scales. 
Unlike the case of ultralight axionlike pseudoscalar particles with negligible (attractive) self-interactions 
($a_s=0$), which necessarily predict solitons at the centres of the haloes with a radius and density that scales 
as $M^{-1}$ and $M^4$, respectively, a small repulsive self-interaction induces a natural minimum scale length 
for the size of (non-linear) dark matter structures and favours central superfluid dark matter droplets that 
lie in either the Thomas--Fermi or the crossover regimes, having an essentially constant radius 
$r_c\,\sim\,1$\,kpc and a shallow density that scales as $M$. This can help solve the small-scale problems of 
the CDM model.
Note that since the theoretical lower limit for $M$ is of the order of the Jeans mass 
$M_\textsc{J} \approx 1 \times 10^7\,\mathrm{M}_{\sun}$, 
there is still room in the self-interacting model for the 
existence of quasi-solitons with maximum radii smaller than $r_c(M_\textsc{J}) \approx 7$\,kpc.

Since for distances much larger than $\lambda_{dB}\,\sim\,1$\,kpc both solitons and TF-like droplets behave 
essentially like CDM particles with mass $\rho \lambda_{dB}^3$ \citep{Schive2014a,Hui2017}, on large scales 
the self-interacting BEC dark matter model becomes virtually indistinguishable from the standard CDM model.
Therefore, in order to corroborate 
our results it would be interesting to probe the cores of the most dark matter dominated galaxies with higher 
resolution, as well as to carry out numerical simulations to check the model beyond the inner part of the 
haloes (where an analytical treatment is applicable) or in the presence of a non-negligible baryonic 
contribution. We hope that our results will stimulate further investigations in these directions.

\section*{Acknowledgements}

We thank J. E. Betancort-Rijo and C. Dalla Vecchia for stimulating discussions. V. D. acknowledges 
support from Agencia Estatal de Investigaci{\'o}n (Ministerio de Ciencia e Innovaci{\'o}n, Spain) and
Fondo Europeo de Desarrollo Regional (FEDER, EU) under Grants PID2019-105225GB-100 and FIS2016-79596-P.

\section*{Data Availability}
The data underlying this article will be shared on reasonable request to the corresponding author.



\bibliographystyle{mnras}
\bibliography{References-v4} 




\appendix

\section{Fitting criteria}
\label{sec:appenA}

In this Appendix we will examine in more detail the criteria we have
followed in our fitting strategy.

\begin{figure}
\begin{center}
\includegraphics[width=0.9\columnwidth]{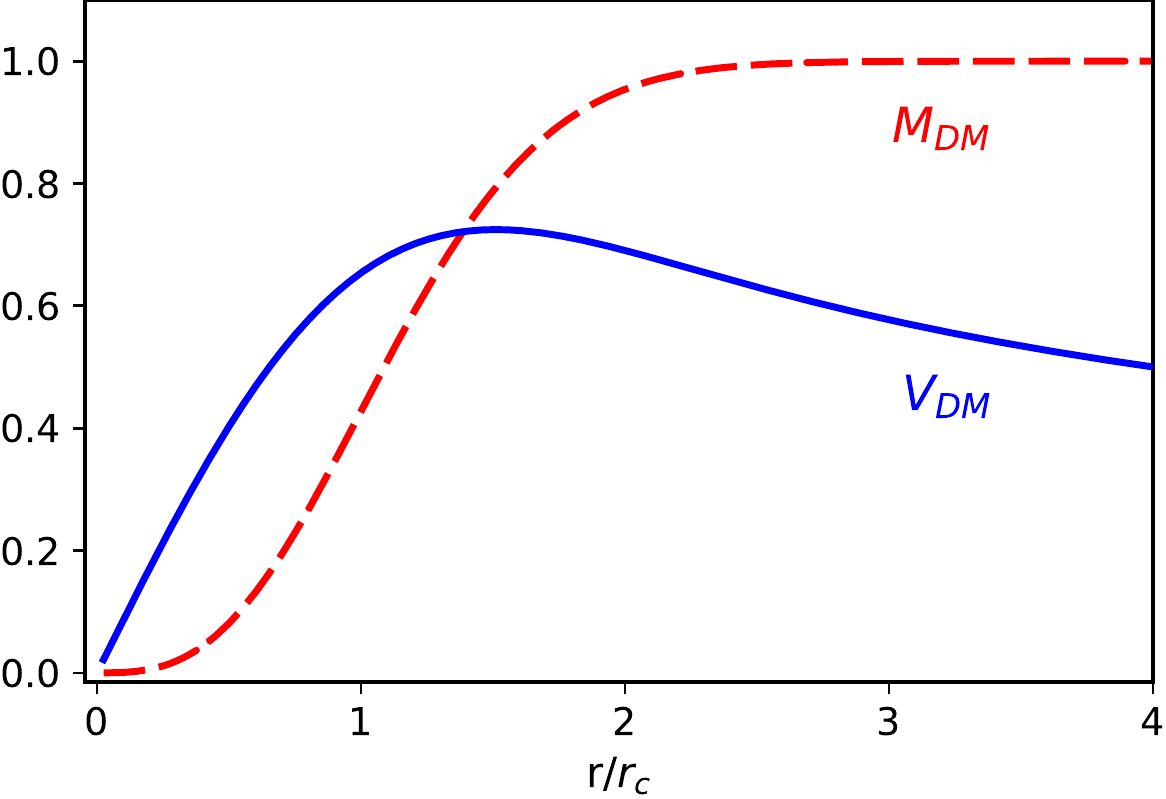}
\end{center}
\caption{Predicted universal mass profile~(\ref{unimass}) of a dark matter 
galactic core (dashed red line) along with the corresponding universal 
rotation curve~(\ref{velo}) (solid blue line). }%
\label{FigA1}%
\end{figure}

Figure~\ref{FigA1} shows the predicted universal mass profile~(\ref{unimass}) 
of a dark matter galactic core along with the corresponding universal 
rotation curve~(\ref{velo}). At $x \equiv r/r_c = 1$, where $r_c$ 
is the radius of the dark matter core given by equation~(\ref{Gaussian_width}), 
the enclosed mass represents $43\%$ of the total mass and the circular 
velocity takes the value $V_{\textsc{dm}}=0.65$.
As can be seen in the figure, the rotation curve attains its maximum, 
$V_{\textsc{dm}}^{\textrm{max}}=0.72$, at $x = 1.5$, and the enclosed mass 
within this radius corresponds to $79\%$ of the total mass.


Consider a strongly dark matter dominated galaxy and assume that, as predicted 
by our model, its observational rotation curve can be well described in the 
innermost region of the galactic halo by the universal profile~(\ref{velo}), up to 
a certain radius $r_{\textsc{l}} \approx \alpha r_c$, where $r_c$ is the radius of the 
central dark matter droplet and $\alpha$ is an unknown factor of order unity.
Let us assume for the moment (we will remove this assumption later) that the 
observational data have negligible uncertainties (error bars).
Since the theoretical curve~(\ref{velo}) only depends on three independent parameters, 
it is sufficient to impose that it passes through any three points \textit{within} the 
radius $r_{\textsc{l}}$ to properly account for the whole curve. 
Thus, in this hypothetical case of accurate observational data, the contribution of 
any other points, as long as they lie within the radius $r_{\textsc{l}}$, should not 
lead to an appreciable modification of the resulting best-fit parameters of the model.
In contrast, by forcing the theoretical curve to pass through points where the model 
is no longer applicable ($r>r_{\textsc{l}}$), a significant bias in the inferred parameters 
can be introduced, and this is true regardless of whether or not the observational data 
have negligible uncertainties.
It is therefore important to prevent observational data beyond the range of 
applicability of the model from intervening in the fitting procedure.
To achieve this, in principle, it would be sufficient to limit the fit to a maximum 
radius $r_{\textsc{f}}<r_{\textsc{l}}$. 

However, while the assumptions of the model allow us to ensure that its range of 
applicability ($r \lesssim \alpha r_c$ with $ \alpha \sim O(1)$) is of the order 
of the radius $r_c$ of the central dark matter droplet, the value of $r_c$ itself 
is not known a priori. In fact, this value is one of the results of the fit.
It is thus clear that some assumption has to be made in order to proceed further. 
In this work, we assume that the model can be applied to a good approximation 
\textit{at least} up to a certain radius $r_{\textsc{f}}$.
In view of the considerations above, the safest and most conservative option to 
avoid biased results is to take the smallest possible value for $r_{\textsc{f}}$.
However, in doing so, one must be aware that the uncertainties in the observational 
data are by no means negligible. 
In fact, it can be readily verified that due to these uncertainties the fits of two
individual galaxies, even within their innermost regions, can lead to quite different
model parameters.
The easiest way to overcome this difficulty is to increase the size of the sample.
As the sample size increases, random uncertainties are expected to cancel out to 
a greater extent, thus increasing our confidence in the observational data and, 
hence, in the model predictions.

One way to increase the sample size would be to choose a larger fit radius 
$r_{\textsc{f}}$, however, as mentioned above, this must be done with care 
to minimise the risk of biased results. 
Observations and numerical simulations seem to indicate that a reasonable value 
would be $r_{\textsc{f}}\sim 1$\,kpc. 
Since most galaxies are poorly resolved in this region,
this is far from enough to compensate for observational uncertainties.
To further increase the sample size, we have taken advantage of the fact that
the model predicts a universal rotation curve in the innermost region of
galactic haloes, which allows us to simultaneously fit all 17 galaxies 
as if they were a single data set.
This is a crucial point in our approach.
Thus, our fitting strategy consists of reducing the impact of observational 
uncertainties by increasing the statistical significance of the sample while 
minimising the risk of biased results. 
To this end, we have simultaneously fit all galaxies and have taken 
the smallest possible fit radius $r_{\textsc{f}}$ that still allows sufficient 
resolution in the region of interest. 
We believe that the best compromise is achieved for $r_{\textsc{f}}=1.2$\,kpc.
The best-fit parameters obtained with this value are collected in Table~\ref{Tab1}.
In particular, the boson mass and the s-wave scattering length are, respectively,
$m \simeq 2.2 \times 10^{-22}$\,eV\,c$^{-2}$ and $a_s \simeq 7.8 \times 10^{-77}$\,m,
while the radii $r_c$ of the central dark matter droplets are all of the order of
$1$\,kpc.

We have also fitted the observational data using $r_{\textsc{f}}=1$ 
and $1.4$\,kpc. In the first case, we have obtained 
$m \simeq 1.8 \times 10^{-22}$\,eV\,c$^{-2}$ and $a_s \simeq 2.5 \times 10^{-77}$\,m.
While this result is in reasonable agreement with the previous one, in this case 
the number of data points to fit is clearly smaller as some of the galaxies that 
previously contributed 6 points now contribute only 5, which makes it advisable 
to now limit the maximum number of points to fit per galaxy to 5 to avoid overweighting 
some galaxies too much. 
On the other hand, for $r_{\textsc{f}}=1.4$\,kpc we have obtained
$m \simeq 1.5 \times 10^{-22}$\,eV\,c$^{-2}$ and $a_s \simeq 3.3 \times 10^{-77}$\,m,
a result which, although in our view obtained with an unnecessarily large value of 
$r_{\textsc{f}}$, is still in reasonably good agreement with the previous results.



\begin{table*}
\caption{Best-fit parameters ($m$, $a_s$, $\{ M \}$). Also shown are the radii $r_c$ and masses $M_c$ of the halo 
cores and the enclosed masses within $r=300$\,pc ($M_{300}$) and corresponding mean densities $\rho_{300}$. } 
\begin{tabular}{lcrccc}
\hline
\multicolumn{6}{c}{$m = 2.2 \times 10^{-22}$\,eV\,c$^{-2} \qquad \qquad a_s = 7.8 \times 10^{-77}$\,m}\\ 
\hline
Galaxy	& $M \; (10^9\,\mathrm{M}_{\sun})$ & $r_c \; (\mathrm{pc})$ & $M_c \; (10^8\,\mathrm{M}_{\sun})$ %
& $M_{300} \; (10^7\,\mathrm{M}_{\sun})$ & $\rho_{300} \; (10^{-1}\,\mathrm{M}_{\sun}\,\mathrm{pc}^{-3})$ \\
\hline 
 UGCA444		&	0.24	&  997{\hspace*{6pt}}	&	1.03  &  0.47  &  0.41  \\
 NGC0024		&	1.89	&  861{\hspace*{6pt}}	&	8.09  &  5.60  &  4.95  \\
 UGC01281		&	0.19	& 1045{\hspace*{6pt}}	&	0.80  &	 0.32  &  0.28  \\
 UGC05721		&	1.31	&  869{\hspace*{6pt}}	&	5.62  &	 3.79  &  3.35  \\
 NGC3741		&	0.31	&  962{\hspace*{6pt}}	&	1.30  &	 0.66  &  0.58  \\
 NGC3109		&	0.17	& 1066{\hspace*{6pt}}	&	0.73  &	 0.27  &  0.24  \\
 NGC6789		&	1.78	&  862{\hspace*{6pt}}	&	7.62  &	 5.26  &  4.65  \\
 UGC07603		&	0.51	&  913{\hspace*{6pt}}	&	2.16  &	 1.27  &  1.12  \\
 KK98--251		&	0.12	& 1189{\hspace*{6pt}}	&	0.49  &	 0.13  &  0.12  \\
 UGC07524		&	0.31	&  959{\hspace*{6pt}}	&	1.34  &	 0.68  &  0.60  \\
 UGC07866		&	0.24	&  997{\hspace*{6pt}}	&	1.02  &	 0.46  &  0.41  \\
 ESO444--G084	&	0.69	&  894{\hspace*{6pt}}	&	2.95  &	 1.84  &  1.63  \\
 UGC05764		&	0.65	&  897{\hspace*{6pt}}	&	2.76  &	 1.70  &  1.50  \\
 F5632			&	0.52	&  911{\hspace*{6pt}}	&	2.23  &	 1.31  &  1.16  \\
 F5834			&	0.33	&  952{\hspace*{6pt}}	&	1.42  &	 0.74  &  0.65  \\
 DDO154			&	0.19	& 1046{\hspace*{6pt}}	&	0.80  &	 0.31  &  0.28  \\
 UGC07232		&	0.72	&  892{\hspace*{6pt}}	&	3.07  &	 1.92  &  1.70  \\ 
\hline 
\end{tabular}
\label{Tab1}
\end{table*}



\bsp	
\label{lastpage}
\end{document}